\documentclass[aps,preprint]{revtex4}%
\usepackage{amsfonts}
\usepackage{amsmath}
\usepackage{amssymb}
\usepackage{graphicx}%
\begin{document}
\title[THERMODYNAMICS OF HARMONIC OSCILLATOR]{THERMODYNAMICS OF THE HARMONIC OSCILLATOR: WIEN'S DISPLACEMENT LAW AND THE
PLANCK SPECTRUM}
\author{Timothy H. Boyer}
\affiliation{Department of Physics, City College of the City University of New York, New
York, NY 10031}

\begin{abstract}
A thermodynamic analysis of the harmonic oscillator is presented. \ Motivation
for the study is provided by the blackbody radiation spectrum; when blackbody
radiation is regarded as a system of noninteracting harmonic oscillator modes,
the thermodynamics follows from that of the harmonic oscillators. \ Using the
behavior of a harmonic oscillator thermodynamic system under an adiabatic
change of oscillator frequency $\omega$, we show that the thermodynamic
functions can all be derived from a single function of $\omega/T$, analogous
to Wien's displacement theorem. The high- and low-frequency energy limits
allow asymptotic energy forms involving $T$ alone or $\omega$ alone,
corresponding to energy equipartition and zero-point energy. It is suggested
that the "smoothest possible" function which behaves as $a-bx$ at small values
of $x$ and vanishes at large $x$ is $ae^{-bx/a}$ because it is a monotonic
analytic function every derivative of which is a multiple of the function
itself. \ In this sense, it is noted that the Planck spectrum with zero-point
radiation corresponds to the function satisfying the Wien displacement result
which provides the smoothest possible interpolation between energy
equipartition at low frequency and zero-point energy at high frequency.

\end{abstract}
\maketitle

\bigskip\draft

\section{\noindent\textbf{INTRODUCTION}}

Blackbody radiation holds an unusual place in the physics curriculum.
\ Blackbody radiation provides a simple system which can be used as a model
for thermodynamic analysis showing the derivation of the Stefan-Boltzmann law
connecting total radiation energy and temperature[1]. \ Also, thermal
radiation can be regarded as a collection of harmonic radiation modes[2] which
can be treated by Doppler shift from a moving piston to obtain the Wien
displacement law relating to the blackbody spectrum[3]. However, the full
blackbody spectrum \ is regarded as inaccessible from thermodynamics; rather
the blackbody spectrum reappears in statistical mechanics where it is used as
an illustration of the breakdown of classical statistical mechanics and the
need for energy quanta[4].

In this article we depart from the traditional points of view. \ We treat
blackbody radiation by analyzing the thermodynamics of \ the harmonic
oscillator. \ Although the harmonic oscillator is often discussed in
statistical mechanics, it is not usually analyzed in thermodynamics[5]. \ We
begin by noting the expression for the work done by a harmonic oscillator on
slow change of its frequency. \ Then we use the laws of thermodynamics to give
a simple but unfamiliar derivation of Wien's displacement law[6]. \ This law
is thus seen to hold not only for electromagnetic radiation but also for any
harmonic oscillator which is a thermodynamic system at temperature $T$. \ Next
we discuss all the thermodynamic functions associated with a harmonic
oscillator thermodynamic system. \ We note that the high- and low-temperature
limits allowed by the Wien law correspond to zero-point energy and energy
equipartition. \ This simple observation is again not traditional in the
physics literature. \ Finally, we suggest a natural interpolation between the
high- and low-temperature forms allowed by the Wien displacement law. \ The
simple interpolation leads to the Planck spectrum with zero-point energy. \ 

The unconventional treatment presented here provides a compact approach to the
blackbody problem through an elementary thermodynamic system which is simple
enough for the textbooks. \ The analysis emphasizes the intimate connections
between blackbody radiation and zero-point radiation, and also fits neatly
with the modern realization that constant acceleration of a harmonic
oscillator through zero-point radiation makes the oscillator energy equal to
the Planck spectrum with zero-point radiation[7].

\section{\noindent\ THERMODYNAMICS OF A HARMONIC OSCILLATOR}

A harmonic oscillator corresponds to a nonrelativistic point mass $m$ moving
in one dimension in a potential $V=(1/2)m\omega^{2}x^{2}$. \ It can be
described by the Lagrangian $L(x,\dot{x})=(1/2)m\dot{x}^{2}-(1/2)m\omega
^{2}x^{2}$, and corresponds to harmonic oscillation $x=(2\mathcal{U}%
/m\omega\mathcal{)}^{1/2}\cos(\omega t+\phi)$ at energy $\mathcal{U}$ and
angular frequency $\omega$. \ Under a slow change in the spring constant of
the system corresponding to a change in the natural angular frequency $\omega
$, the energy divided by the frequency $\mathcal{U}/\omega$ is a constant[8]
and so the change in system energy is $d\mathcal{U}=(\mathcal{U}%
/\omega)d\omega$ and the work $dW$ done by the system on the external agency
changing the frequency is%
\begin{equation}
dW=-(\mathcal{U}/\omega)d\omega.
\end{equation}
Thus we regard $\mathcal{X}=-\mathcal{U}/\omega$ as the generalized force
associated with a change in frequency $\omega$.

If a harmonic oscillator is weakly coupled to a heat bath at temperature $T$,
then it will exchange energy with the heat bath. \ In equilibrium with the
bath, the harmonic oscillator at frequency $\omega$ will have an average
energy $\mathcal{U}(\omega.T)$. \ Thus the oscillator forms a thermodynamic
system suitable for treatment by the usual methods of thermodynamics involving
(average) energy $\mathcal{U}$, parameter $\omega$, and associated work
$dW=-(\mathcal{U}/\omega)d\omega$.

\section{\noindent\textbf{DERIVATION OF FUNCTIONAL DEPENDENCE CORRESPONDING TO
WIEN'S DISPLACEMENT THEOREM}}

When the oscillator is at thermal equilibrium in a bath at temperature $T$,
the oscillator will have an (average) energy $\mathcal{U}(\omega,T)$ and an
entropy $\mathcal{S}(\omega,T)$ depending upon the temperature $T$ and natural
oscillation frequency $\omega$. Using the work expression in(1), the laws of
thermodynamics for the oscillator give
\begin{equation}
dQ=Td\mathcal{S}(T,\omega)=d\mathcal{U}(T,\omega)-(\mathcal{U}/\omega)d\omega
\end{equation}
This can be rewritten as
\begin{equation}
T\left[  \left(  \frac{\partial\mathcal{S}}{\partial T}\right)  _{\omega
}dT+\left(  \frac{\partial\mathcal{S}}{\partial\omega}\right)  _{T}%
d\omega\right]  =\left(  \frac{\partial\mathcal{U}}{\partial T}\right)
_{\omega}dT+\left(  \frac{\partial\mathcal{U}}{\partial\omega}\right)
_{T}d\omega-\frac{\mathcal{U}}{\omega}d\omega
\end{equation}
Since the variables $\omega$ and $T$ are independent, this requires
\begin{equation}
T\left(  \frac{\partial\mathcal{S}}{\partial T}\right)  _{\omega}=\left(
\frac{\partial\mathcal{U}}{\partial T}\right)  _{\omega},~~~~~T\left(
\frac{\partial\mathcal{S}}{\partial\omega}\right)  _{T}=\left(  \frac
{\partial\mathcal{U}}{\partial\omega}\right)  _{T}-\frac{\mathcal{U}}{\omega}%
\end{equation}
Differentiating the first equation of (4) with respect to $\omega$ and the
second with respect to $T$ gives
\begin{equation}
T\frac{\partial^{2}\mathcal{S}}{\partial\omega\partial T}=\frac{\partial
^{2}\mathcal{U}}{\partial\omega\partial T},~~~~~\left(  \frac{\partial
\mathcal{S}}{\partial\omega}\right)  _{T}+T\frac{\partial^{2}\mathcal{S}%
}{\partial T\partial\omega}=\frac{\partial^{2}\mathcal{U}}{\partial
T\partial\omega}-\frac{1}{\omega}\left(  \frac{\partial\mathcal{U}}{\partial
T}\right)  _{\omega}%
\end{equation}
Subtracting the first equation of (5) from the second as as to eliminate the
second-derivative terms, we have
\begin{equation}
\left(  \frac{\partial\mathcal{S}}{\partial\omega}\right)  _{T}=-\frac
{1}{\omega}\left(  \frac{\partial\mathcal{U}}{\partial T}\right)  _{\omega}%
\end{equation}
Next using (6) together with the first equation of (4), we find
\begin{equation}
\left(  \frac{\partial\mathcal{S}}{\partial\omega}\right)  _{T}=-\frac
{T}{\omega}\left(  \frac{\partial\mathcal{S}}{\partial T}\right)  _{\omega}%
\end{equation}
which has the general solution
\begin{equation}
\mathcal{S}(\omega,T)=g(\omega/T)
\end{equation}
where $g$ is an arbitrary function of the single variable $\omega/T$. On the
other hand if we use the second equation of (4), then we find from Eq.(6)
\begin{equation}
\left(  \frac{\partial\mathcal{U}}{\partial\omega}\right)  _{T}-\frac
{\mathcal{U}}{\omega}=-\frac{T}{\omega}\left(  \frac{\partial\mathcal{U}%
}{\partial T}\right)  _{\omega}%
\end{equation}
which has the general solution
\begin{equation}
\mathcal{U}(\omega,T)=\omega f(\omega/T)
\end{equation}
where $f$ is an arbitrary function of the single variable $\omega/T$. This
corresponds to the information in the Wien displacement theorem.[3] Although
the Wien theorem is often derived from the doppler shift of electromagnetic
waves reflected from a moving piston in a cavity, our analysis shows that it
holds in thermal equilibrium for any system described by a simple harmonic
oscillator Lagrangian.

\section{\noindent\textbf{THERMODYNAMIC FUNCTIONS FOR A HARMONIC OSCILLATOR}}

The Wien displacement results\ $\mathcal{S}(\omega,T)=g(\omega/T)$ and
$\mathcal{U}(\omega,T)=\omega f(\omega/T)$ give constraints upon all the
thermodynamic functions for a harmonic oscillator. The thermodynamic
potential[9] $\phi(\omega/T)$, from which all the other thermodynamics
functions may be derived, must be a function of the combination $\omega/T$.
The average oscillator energy $\mathcal{U}$ in thermal equilibrium follows as
\begin{equation}
\mathcal{U}(\omega,T)=T^{2}\left(  \frac{\partial\phi}{\partial T}\right)
_{\omega}=-\omega\phi^{\prime}(\omega/T)
\end{equation}
The entropy $\mathcal{S}$ of the oscillator is again a function of $\omega
/T$,
\begin{equation}
\mathcal{S}(\omega/T)=\phi(\omega/T)+\mathcal{U}(\omega,T)/T=\phi
(\omega/T)-(\omega/T)\phi^{\prime}(\omega/T)
\end{equation}
The Helmholtz free energy $\mathcal{F}$ is directly related to the
thermodynamic potential $\phi(\omega/T)$
\begin{equation}
\mathcal{F}(\omega,T)=-T\phi(\omega/T)
\end{equation}
The generalized force $\mathcal{X}$ associated with a change in $\omega$ is
\begin{equation}
\mathcal{X}(\omega/T)=T\left(  \frac{\partial\phi}{\partial\omega}\right)
_{T}=\phi^{\prime}(\omega/T)
\end{equation}
and the specific heat $\mathcal{C}$ at constant $\omega$ is given by
\begin{equation}
\mathcal{C}(\omega/T)=\left(  \frac{\partial\mathcal{U}}{\partial T}\right)
_{\omega}=\left(  \frac{\omega}{T}\right)  ^{2}\phi^{\prime\prime}(\omega/T)
\end{equation}
Thus the equilibrium thermodynamics of a classical harmonic oscillator system
is determined by one function, the unknown function $\phi(\omega/T)$.

\section{\noindent\textbf{WIEN DISPLACEMENT RESULT AND ZERO-POINT ENERGY}}

There are two natural extremes for the oscillator energy given by the Wien
displacement result in (11); one extreme makes the energy $\mathcal{U}%
(\omega,T)$ independent of temperature $T$, and the other makes the energy
$\mathcal{U}(\omega,T)$ independent of the natural frequency $\omega$.

We deal first with the temperature-independent energy. When the potential
function $\phi^{\prime}(\omega/T)=-const$ and so
\begin{equation}
\phi(\omega/T)=-const\times(\omega/T)
\end{equation}
then the oscillator energy in (11) takes the form
\begin{equation}
\mathcal{U}(\omega,T)=\mathcal{U}_{zp}(\omega)=const\times\omega
\end{equation}
This corresponds to temperature-independent zero-point energy.

We note that substitution of the zero-point energy (17) into the first law of
thermodynamics in the form
\begin{equation}
dQ=d\mathcal{U}-(\mathcal{U}/\omega)d\omega=d(const\times\omega)-(const\times
\omega/\omega)d\omega=0
\end{equation}
tells us that no heat $dQ$ enters the system on changing the natural frequency
of the oscillator $\omega$. Thus changes in zero-point energy occur without
any change in the thermodynamic entropy $\mathcal{S}(\omega/T)$ of the system.
Indeed, we see that if $\phi_{zp}^{\prime}$ is constant, then $\phi
_{zp}(\omega/T)=const\times\omega/T$ must be linear in its argument, and the
entropy $\mathcal{S}$ in Eq.(12) vanishes for any function $\phi$ which is
linear in its argument.

Zero-point energy is random energy which is present even at zero temperature.
Thermodynamics allows the possibility of zero-point energy and experimental
evidence (such as that for van der Waals forces) requires its existence.[10]
It is natural to choose the unknown constant for the zero-point energy in (17)
so as to fit the experimentally measured van der Waals forces. This
corresponds to an oscillator energy
\begin{equation}
\mathcal{U}_{zp}(\omega)=(1/2)\hbar\omega
\end{equation}
where $\hbar$ is a constant which takes the value familiar for Planck's constant.

\section{\noindent\textbf{WIEN DISPLACEMENT RESULT AND ENERGY EQUIPARTITION}}

The other extreme for the Wien displacement result (11) is the case where the
oscillator energy depends upon the temperature but has no dependence upon the
natural oscillator frequency $\omega$. Thus when $\phi^{\prime}(\omega
/T)=-const/(\omega/T)$ in equation (11), then the oscillator energy is
\begin{equation}
\mathcal{U}(\omega,T)=\mathcal{U}_{RJ}(T)=const\times T.
\end{equation}
This is the familiar energy equipartition law (proposed by Rayleigh and Jeans
for low-frequency radiation modes) where the constant is chosen as Boltzmann's
constant $k_{B}$,
\begin{equation}
\mathcal{U}_{RJ}(T)=k_{B}T.
\end{equation}

In this case, an isothermal change of the natural oscillator frequency
$\omega$ produces no change in the oscillator internal energy. Rather, from
(2), the isothermal work done on changing the natural frequency $\omega$ is
provided by heat added which keeps the internal oscillator energy constant,
\begin{equation}
dQ=Td\mathcal{S}(\omega/T)=d\mathcal{U}_{RJ}(T)-(\mathcal{U}/\omega
)d\omega,~~~constant\text{ }T
\end{equation}
Then
\begin{equation}
d\mathcal{S}_{RJ}=0-(k_{B}/\omega)d\omega,~~~constant\text{ }T,
\end{equation}
and since we know the functional form $\mathcal{S}(\omega/T)$, we have the
familiar result
\begin{equation}
\mathcal{S}_{RJ}(\omega/T)=-k_{B}\ln(\omega/T)+const
\end{equation}
Indeed if $\phi_{RJ}^{\prime}(\omega/T)=-k_{B}/(\omega/T)$, then $\phi
_{RJ}(\omega/T)=-k_{B}\ln(\omega/T)$ and the entropy in (12) takes the form (24).

\section{\noindent\textbf{USE OF NATURAL UNITS IN THE ANALYSIS}}

In this paper we are not interested in the numerical evaluation of
thermodynamic quantities but rather in the fundamental thermodynamic behavior.
On this account we will measure all quantities in terms of energy and take the
entropy as a pure number. Thus we will take $\hbar=1$ and measure frequencies
in energy units. Also, we will take $k_{B}=1$ and measure temperature in
energy units.[11] \ Thus the limiting form corresponding to zero-point energy
has
\begin{equation}
\phi_{zp}(\omega/T)=-(1/2)(\omega/T),~~~~~\mathcal{U}_{zp}(\omega)=(1/2)\omega
\end{equation}
while the limiting form corresponding to energy equipartition becomes
\begin{equation}
\phi_{RJ}(\omega/T)=-\ln(\omega/T),~~~~~\mathcal{U}_{RJ}(T)=T
\end{equation}

\noindent

\section{\noindent\textbf{ASYMPTOTIC LIMITS FOR THERMAL OSCILLATOR ENERGY}}

In general, the behavior of an oscillator system will depend upon both
frequency $\omega$ and temperature $T$ as in (11). In the limit as
$T\rightarrow0$, we expect to recover the zero-point energy of the oscillator
\begin{equation}
lim_{T\rightarrow0}\mathcal{U}(\omega,T)=lim_{T\rightarrow0}[-\omega
\phi^{\prime}(\omega/T)]=\mathcal{U}_{zp}(\omega)=(1/2)\omega
\end{equation}
and the associated thermodynamic potential%
\begin{equation}
\phi(\omega/T)\rightarrow\phi_{zp}(\omega/T)=-(1/2)(\omega/T)
\end{equation}
In the limit $\omega\rightarrow0$, we expect to obtain the equipartition
energy
\begin{equation}
lim_{\omega\rightarrow0}\mathcal{U}(\omega,T)=lim_{\omega\rightarrow0}%
[-\omega\phi^{\prime}(\omega/T)]=\mathcal{U}_{RJ}(T)=T
\end{equation}
and the associated thermodynamic potential%
\begin{equation}
\phi(\omega/T)\rightarrow\phi_{RJ}(\omega/T)=-\ln(\omega/T)
\end{equation}

It is sometimes useful to make a distinction between the THERMAL energy
$\mathcal{U}_{T}(\omega,T)$ of an oscillator and the oscillator's TOTAL energy
$\mathcal{U}(\omega,T)$. The thermal energy is just the (average) energy above
the (average) zero-point energy
\begin{equation}
\mathcal{U}_{T}(\omega,T)=\mathcal{U}(\omega,T)-\mathcal{U}_{zp}%
(\omega)=-\omega\phi^{\prime}(\omega/T)-(1/2)\omega=-\omega\lbrack\phi
^{\prime}(\omega/T)-\phi_{zp}^{\prime}(\omega/T)]
\end{equation}
Although the total oscillator energy $\mathcal{U}(\omega,T)$ is related to
forces, it is only the thermal oscillator energy $\mathcal{U}_{T}(\omega,T)$
which is related to changes in thermodynamic entropy since (as seen above)
$\phi_{zp}(\omega/T)$ does not give any thermodynamic entropy. \ 

\section{\noindent\textbf{PLANCK SPECTRUM AS THE SMOOTHEST INTERPOLATION
BETWEEN THE EQUIPARTITION AND ZERO-POINT LIMITS}}

Although the thermodynamic forms given in Eqs.(11)-(15) represent the
information obtained from thermodynamic analysis, it is tempting to try to
guess the full thermodynamic behavior which is chosen by nature. \ Now
experiments indicate that nature indeed chooses the asymptotic forms given in
Eqs.(27)-(30) corresponding to non-zero values of $\hbar$ and $k_{B\text{.}}$
\ Thus we ask whether there is any natural choice for behavior which connects
these asymptotic forms. \ One expects thermodynamics to involve smooth
functions, and hence one might seek the "smoothest possible" interpolation
between the extremes.

Is the idea of a "smoothest possible" interpolation well-defined? \ In
general, this seems ambiguous. \ However, there are cases where the "smoothest
possible" function appears absolutely clear. \ Thus if for small values of $x$
an analytic function has the form $f(x)=a-bx+O(x^{2})$ with $a>0$, $b>0$, and
for large values of $x$ the function $f(x)\rightarrow0,$ then the smoothest
possible interpolation between the limits is $f(x)=a\exp(-bx/a)$. \ This
function meets the asymptotic limits and is a monotonic function whose $n$th
derivative is just $(-b/a)^{n}$ times the value of the function at that point.
\ Thus every point $x$ involves the same local functional behavior for $f(x)$.
\ Every point is equivalent. \ It is only modification of this functional form
which introduces preferred values for $x$.

Let us reconsider the asymptotic forms given in Eqs.(27)-(30). \ The energy
forms are not useful since these do not give us asymptotic functions of
$\omega/T$. \ The limiting thermodynamic potentials
\begin{equation}
\phi_{RJ}(\omega/T)=-\ln(\omega/T),~~~~~\phi_{zp}(\omega/T)=-(1/2)\omega/T
\end{equation}
involve logarithmic behavior at the low-frequency limit. Since the logarithmic
function is more complicated analytically than the exponential function, it is
convenient to take the exponential of the negative of these functions and to
consider
\begin{equation}
\exp[-\phi_{RJ}(\omega/T)]=\omega/T,~~~\exp[-\phi_{zp}(\omega/T)]=\exp
[\omega/(2T)]
\end{equation}
The exponentiation will not change the "smoothest possible" criterion required
of the interpolation. \ Thus according to (33) we are searching for the
"smoothest possible" interpolation $\exp[-\phi(\omega/T)]$ between linear
behavior $\omega/T$ at small argument and exponential behavior $\exp
[\omega/(2T)]$\ at large argument%
\begin{equation}
\exp[-\phi(\omega/T)]\rightarrow\omega/T\text{ for }\omega/T\rightarrow
0\text{, \ and \ }\exp[-\phi(\omega/T)]\rightarrow\exp[\omega/(2T)]\text{ for
}\omega/T\rightarrow\infty
\end{equation}
We notice that the difference of the high-frequency limit and the desired
interpolation has the asymptotic forms
\[
\exp[\omega/(2T)]-\exp[-\phi(\omega/T)]\rightarrow1+\omega/(2T)-\omega
/T=1-\omega/(2T)\text{ \ \ for }\omega/T\rightarrow0
\]%
\begin{equation}
\exp[\omega/(2T)]-\exp[-\phi(\omega/T)]\rightarrow0\text{ \ \ for \ }%
\omega/T\rightarrow\infty
\end{equation}
But this corresponds to exactly the case mentioned in the previous paragraph.
\ Thus the "smoothest possible" interpolation is
\begin{equation}
\exp[\omega/(2T)]-\exp[-\phi(\omega/T)]=\exp[-\omega/(2T)]
\end{equation}
This implies%
\begin{equation}
\exp[-\phi(\omega/T)]=\exp[\omega/(2T)]-\exp[-\omega/(2T)]=2\sinh[\omega/(2T)]
\end{equation}

It is easy to see that the right-hand side of (37) has exactly the asymptotic
forms demanded in (33). Then taking the logarithm of (37) and using the
thermodynamic relations given in (11) and (12), this smooth interpolation
leads to the thermodynamic functions
\begin{equation}
\phi_{Pzp}\left(  \frac{\omega}{T}\right)  =-\ln\left[  2\sinh\left(  \frac
{1}{2}\frac{\omega}{T}\right)  \right]
\end{equation}%
\begin{equation}
\mathcal{U}_{Pzp}\left(  \omega,T\right)  =\frac{1}{2}\omega\coth\left(
\frac{1}{2}\frac{\omega}{T}\right)  =\frac{\omega}{\exp(\omega/T)-1}+\frac
{1}{2}\omega
\end{equation}%
\begin{equation}
\mathcal{S}_{P}\left(  \frac{\omega}{T}\right)  =-\ln\left[  2\sinh\left(
\frac{1}{2}\frac{\omega}{T}\right)  \right]  +\frac{1}{2}\frac{\omega}{T}%
\coth\left(  \frac{1}{2}\frac{\omega}{T}\right)
\end{equation}
We have labeled these thermodynamic functions with the subscript "Pzp" or "P"
because they correspond exactly to the familiar Planck average oscillator
energy including zero-point energy. \ As noted above, the entropy depends upon
the Planck thermal spectrum $\mathcal{U}_{T}(\omega,T)$ but does not reflect
the zero-point energy included in $\mathcal{U}(\omega,T)$.

\section{COMMENTS ON THE INTERPOLATION FOR PLANCK'S SPECTRUM}

The analysis above shows that demanding the smoothest interpolation between
the equipartition and zero-point limits suggests the Planck spectrum. \ This
seems a surprisingly simple extrapolation from the thermodynamic analysis.
\ One may wonder why such an extrapolation was not made a century ago.
\ Indeed, many physicists are aware that Planck did arrive at the blackbody
spectrum as an interpolation. \ However, Planck's interpolation did not come
from the limits on the Wien displacement theorem. \ Rather, his interpolation
involved a modification which combined earlier guesses at the high- and
low-frequency parts of the thermal energy spectrum $\mathcal{U}_{T}(\omega,T)$.[12]

The simple interpolation made here was not made a century ago because
physicists did not think in terms of a temperature-independent zero-point
radiation. \ Indeed, even today, textbook discussions of blackbody radiation
do not usually make any reference to zero-point radiation.[13] \ It is only
much more recently, and in particular in connection with experimental
measurements of Casimir forces, that physicists have taken seriously the
zero-point energy suggested by the high-frequency limit of the Wien
displacement law.

Indeed the smooth extrapolation analysis for a harmonic oscillator given here
fits neatly with the thermal effects of acceleration suggested by Davies and
Unruh[7]. \ If one considers a charged harmonic oscillator undergoing constant
acceleration through zero-point radiation, then the oscillator comes into
equilibrium with the random radiation. \ When the oscillator has zero
acceleration, its average energy $\mathcal{U}(\omega,a=0)$ depends only upon
its frequency and is just zero-point energy $\mathcal{U}(\omega,a=0)=(1/2)$
$\hbar\omega$. \ When the oscillator has a large acceleration $a$, then the
average oscillator energy depends only upon its acceleration $a$ and is
independent of its natural frequency $\omega$. \ Thus zero--point energy and
energy equipartition form the natural extremes. \ And the function which
connects these extremes is found to be exactly the Planck spectrum (39) with a
temperature $T=$ $\hbar a/2\pi ck_{B}$.[7]

\section{CONCLUDING SUMMARY}

The harmonic oscillator provides a simple thermodynamics system involving
energy $\mathcal{U}$, temperature $T$, and oscillator frequency $\omega$. \ If
one carries out a quasi-static change in the oscillator frequency $\omega$,
then thermodynamic analysis leads to results which correspond to the Wien
displacement law. \ All of the thermodynamic behavior for the oscillator can
be derived from a single thermodynamic potential function $\phi(\omega/T)$
depending upon the single variable $\omega/T$. \ If we consider the limits
which make the oscillator energy $\mathcal{U}(\omega,T)$ independent of one of
its variables, then we find the extremes corresponding to zero-point energy
and energy equipartition. \ Finally, as a guess for nature's choice of
thermodynamic behavior, we ask for the function satisfying the Wien
displacement law which provides the smoothest interpolation between the
extremes of zero-point energy and energy equipartition. \ This leads to the
Planck spectrum with zero-point radiation.


\begin{thebibliography}{99}
\bibitem {5}I do not know of any text where the harmonic oscillator is treated
as a thermodynamics system apart from statistical mechanics.
\bibitem {6}I am not aware of any thermodynamic treatment of the adiabatic
change of the harmonic oscillator frequency. \ The familiar derivations of the
Wien law take the Doppler-shift form given in references 3.
\bibitem {13}I am not aware of any textbook treatment of blackbody radiation
which mentions zero-point energy as part of the spectrum involved.
\end{thebibliography}
\end{document}